\documentclass[sigconf]{acmart}
\usepackage{multirow}
\AtBeginDocument{%
  }

\setcopyright{acmlicensed}
\copyrightyear{2025}
\acmYear{2025}
\acmDOI{XXXXXXX.XXXXXXX}
\acmConference[Pre-print submitted to VRST '25]{30th ACM Symposium on Virtual Reality Software and Technology}{November 12--14,
  2025}{Montreal, Canada}
\acmISBN{978-1-4503-XXXX-X/2025/06}

\begin{document}

\title[Music Expertise and Cognitive Load in a VR Rhythm Exergame]{Effects of task difficulty and music expertise in VR: Observations of cognitive load and task accuracy in a rhythm exergame}

\author{Kyla Ellahiyoun}
\email{kyla.ellahiyoun@hotmail.com}
\affiliation{%
  \institution{RMIT University}
  \city{Melbourne}
  \country{Australia}
}

\author{Emma Jane Pretty}
\orcid{0000-0002-5108-5740}
\affiliation{%
  \institution{Tampere University}
  \city{Tampere}
  \country{Finland}}
\email{emma.pretty@tuni.fi}

\author{Renan Guarese}
\orcid{0000-0003-1206-5701}
 \email{guarese@kth.se}
 \affiliation{%
   \institution{Digital Futures, KTH}
   \city{Stockholm}
   \country{Sweden}
 }

 \author{Marcel Takac}
\email{marcel.takac@rmit.edu.au}
\orcid{0000-0002-5761-1828}
\affiliation{%
  \institution{RMIT University}
  \city{Melbourne}
  \state{Victoria}
  \country{Australia}
  \postcode{3003}
}

 \author{Haytham Fayek}
\email{haytham.fayek@rmit.edu.au}
\orcid{0000-0002-1840-7605}
\affiliation{%
  \institution{RMIT University}
  \city{Melbourne}
  \state{Victoria}
  \country{Australia}
  \postcode{3003}
}

 \author{Fabio Zambetta}
\email{fabio.zambetta@rmit.edu.au}
\orcid{0000-0003-4133-7913}
\affiliation{%
  \institution{RMIT University}
  \city{Melbourne}
  \state{Victoria}
  \country{Australia}
  \postcode{3003}
}

\renewcommand{\shortauthors}{Anon et al.}

\begin{abstract}
This study explores the relationship between musical training, cognitive load (CL), and task accuracy within the virtual reality (VR) exergame Beat Saber across increasing levels of difficulty. Participants ($N=32$) completed a series of post-task questionnaires after playing the game under three task difficulty levels while having their physiological data measured by an Emotibit. Using regression analyses, we found that task difficulty and gaming experience significantly predicted subjective CL, whereas musical training did not. However, musical training significantly predicted higher task accuracy, along with lower subjective CL, increased gaming experience, and greater physiological arousal. These results suggest that musical training enhances task-specific performance but does not directly reduce subjective CL. Future research should consider alternative methods of grouping musical expertise and the additional predictability of flow and self-efficacy.
\end{abstract}


\begin{teaserfigure}
  \includegraphics[width=0.52\textwidth]{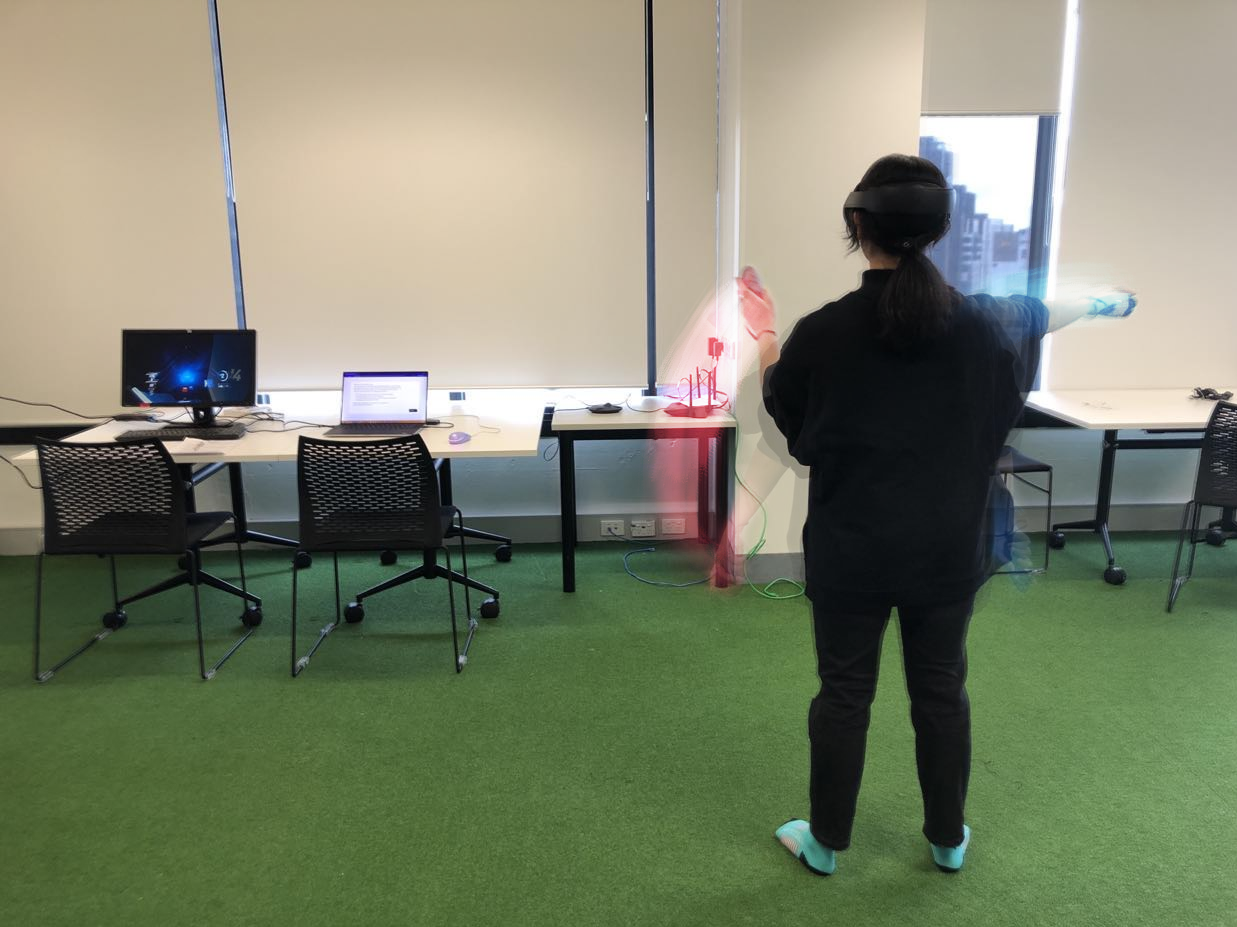}
  \includegraphics[width=0.467\textwidth]{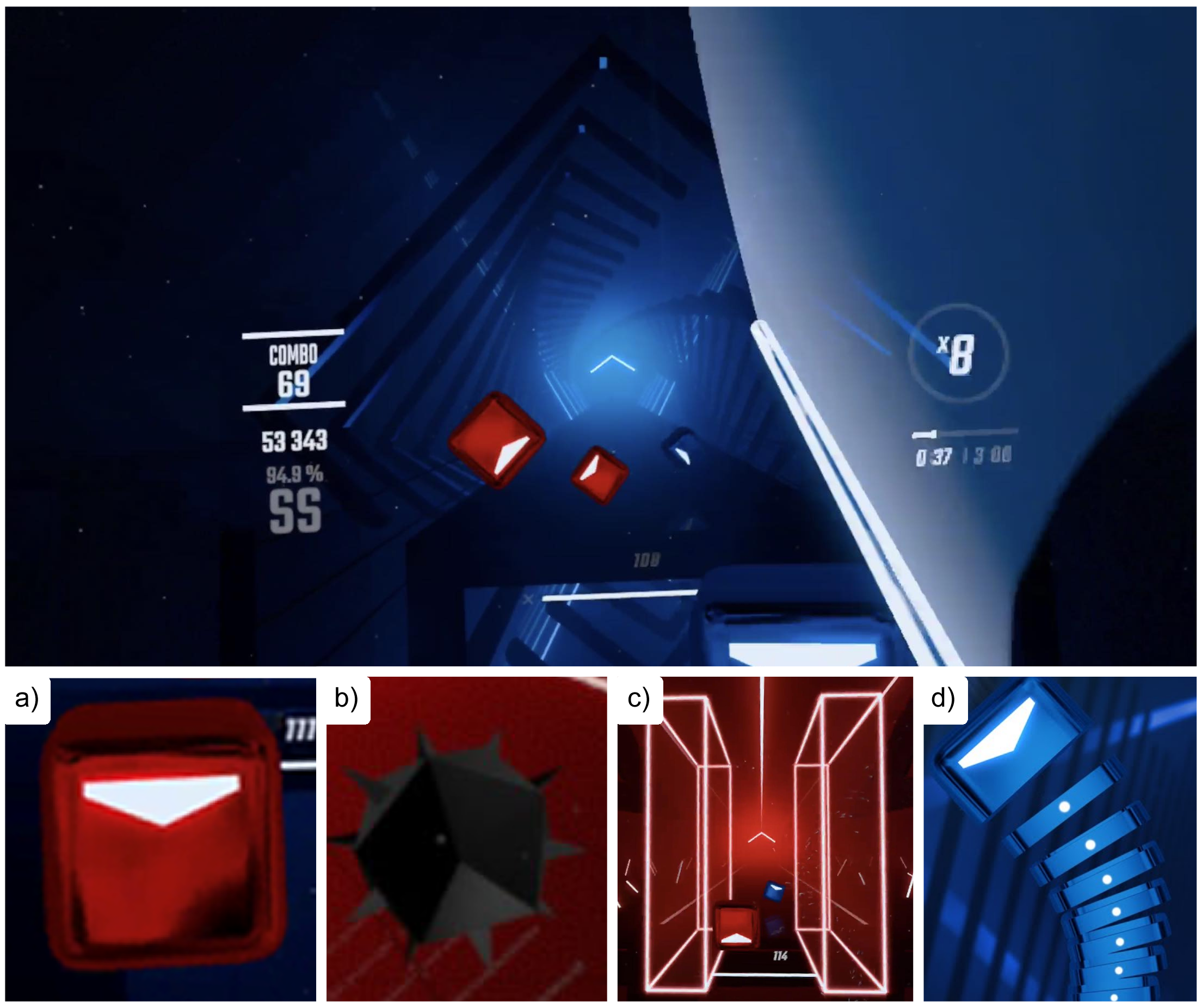}
  \caption{Left -- Superimposed photo of the experimental layout: The screen on the left displays the casting of Beat Saber from the MetaQuest Pro, which is being recorded on OBS. The laptop is used for all surveys, and the participant stands in an open space unobstructed by objects during gameplay. Right -- Screenshots from Beat Saber and its main interactable Objects: a) blocks which represent notes of a song, b) mines which players should avoid, c) walls (an obstacle) that players’ head should avoid, 
  and d) chain notes which represent sounds that are continuous.}
  \Description{Experimental Layout
}
  \label{fig:teaser}
\end{teaserfigure}

\received{09 July 2025}

\maketitle

\section{Introduction}
Music psychology has long been investigating the effect of music on cognitive functions. This has resulted in the realisation that music has positive outcomes on executive functioning \cite{chee_musical_2022}, and that listening behaviour and training could protect against cognitive decline in older populations \cite{roman2018musical}. Recently, this interest in cognitive outcomes is extended to observing cognitive load (CL) during music-oriented tasks. CL accounts for limitations of working memory and considers that the delivery mode of novel information should follow the scientific principle of parsimony \cite{sweller_cognitive_1998}, being presented with the simplest explanation. This is important when presenting new information or tasks, as comparing the expertise between individuals or groups of people can be explained through theories of optimal performance \cite{yerkes1908relation, csikszentmihalyi2000beyond}. In relation to CL and expertise, there is evidence of a negative relationship between a higher level of skill and acquired mental workload during challenging tasks \cite{haith2018multiple}.
CL is prominently researched with regards to instructional design \cite{sweller_cognitive_1998}, where it is established that the inherent difficulty of a task and its delivery method could significantly affect the accumulation of CL \cite{sweller_cognitive_1998}. To address this effect, some researchers have opted for commercial games that are recognisable to the public \cite{huang_exergaming_2020}. A notable example being Beat Saber\footnote{https://www.beatsaber.com/}, a popular exergame \cite{mueller2014movement} that combines upbeat music and physical movement. There have been few attempts to modify games such as Beat Saber or create new rhythm games for research participants to engage with \cite{lemmens_let_2023}, however, they are susceptible to limited generalisation as they can lack elements which make games inherently enjoyable \cite{cutting_difficulty-skill_2023}.

\subsection{Music Sophistication \& Training}

An abundance of research has defined musical abilities by the duration of formal musical training (MT), ignoring other factors that do not overtly contribute to the skillset \cite{levitin2012does, sonification:2022}. A widely adopted measure of musical ability in the general population is the Goldsmiths Music Sophistication Index (Gold-MSI) \cite{mullensiefen_musicality_2014}. It observes music sophistication as a psychometric construct, encompassing independent dimensions comprising skills and achievements, musical perception and music-making, amount of practice, emotional and functional usage of music, and creativity \cite{mullensiefen_musicality_2014}. Despite this, its MT subscale alone is frequently used to group participants, likely due to its perceived objectivity and alignment with more traditional conceptions of musical expertise. For example, studies have chosen certain items from the MT subscale to categorise low, intermediate, and high music sophistication groups \cite{matziorinis_is_2023, audioInterfaces2024}, or only considered MT to inform their definition of expertise \cite{chaddock-heyman_musical_2021, sonification:2022}. As a result, inconsistencies in how musical training is operationalised across studies complicate direct comparison of findings, leading to an overlook of the broader cognitive or experience factors that contribute to task performance.

\subsection{Cognitive Load Theory}

Cognitive Load Theory was conceived because of competing learning and memory frameworks \cite{atkinson1968human, baddeley1974hitch}, demonstrating its adaptability for instructional design \cite{sweller_cognitive_1998}. Furthermore, by accounting for limitations of working memory, the delivery mode of novel information should be parsimonious, following the simplest explanations \cite{sweller_cognitive_1998}. This notion is supported by widely accepted assumptions within CL research, such as element interactivity \cite{sweller1994cognitive, sweller_element_2010} and schema acquisition \cite{chi1981expertise}. Firstly, CL is considered a multidimensional construct, which describes the mental workload that performing a given task imposes on a learner’s cognition \cite{paas1994instructional, sweller_cognitive_1998}. Continuing from this, element interactivity refers to the number of elements being processed simultaneously in working memory for schema construction and their interactions to occur \cite{sweller1994cognitive}. Essentially, it describes the inherent difficulty of a task. Furthermore, schema acquisition relies on long-term memory to guide behaviour from a repository of stored information \cite{kalyuga2016rethinking}. As both of these processes are shaped by prior knowledge and experience, researchers exploring the relationship between CL and task performance increasingly acknowledge the role of domain-specific expertise.

\subsection{Expertise in Context}

Expertise is characterised as a measurable degree of skill, whereby a necessary level of skill or knowledge is objectively referred to as competency \cite{greenfield1994two}. In acknowledgement of this, \citeauthor{kirschner_experts_2013} \cite{kirschner_experts_2013} argued that expertise should reframe the expert versus novice dichotomy by positioning players as experts in their own understanding. It competes with the notion of invalidating the experiences of beginners by focusing on the agency of ‘experts’ \cite{kirschner_experts_2013}. Following their example, \citeauthor{toft-nielsen_gaming_2016} \cite{toft-nielsen_gaming_2016} argued that to acquire expertise in games, the amount of knowledge required to become an expert should be considered. Accordingly, research measuring expertise in virtual reality (VR) exergames should appraise the contextual parameters used to define expertise. In the present study, participants are grouped using the MT subscale of the Gold-MSI, but our focus extends beyond music-specific skill. In addition to musical training, experience levels in digital games, VR, and Beat Saber are included as indicators of broader task-relevant expertise. Taken together, these varying forms of experience may help explain differences in both subjective CL and task accuracy, particularly in interactive, high-paced environments.

In this study, we examine how musical training, gaming experience, and task difficulty interact to influence CL and performance in a VR rhythm exergame. Using Beat Saber as a testbed, participants completed gameplay tasks of increasing difficulty while physiological (electrodermal activity; EDA) and subjective measures of CL (cognitive subscale of the video game demand scale; VGDS \cite{bowman_development_2018}) were collected. This study adds to the growing body of work on individual differences in immersive gameplay experiences, making the following contributions:

    \begin{itemize} 
        \item An investigation on how domain-specific experience and task difficulty impact both subjective CL and performance; 
        \item Evidence that musical training enhances task accuracy in a rhythm exergame environment;
        \item A highlight on the utility of EDA as a physiological predictor of performance in rhythm-based VR contexts. 
    \end{itemize}

The remainder of this paper is structured as follows: Section~\ref{sec:relatedworks} reviews prior work on CL and expertise in games, exergames, and VR. Section~\ref{sec:RQ} defines the research questions guiding this study, while  Section~\ref{sec:methodology} outlines the experimental design, task conditions, and metrics used. Section~\ref{sec:results} presents our key findings, followed by our interpretation of them and their limitations in Section~\ref{sec:discussion}, and finally concluding in Section~\ref{sec:conclusion}.

\section{Related Works}
\label{sec:relatedworks}
As the present study seeks to understand expertise within cognitive engagement, the discussion changes course to consolidate motivational theories into practice \cite{moreno_interactive_2007, huang_motivation_2021}.  

\subsection{Cognitive Load in games}

A popular use of CL is adapting game or character behaviour in real-time using physiological metrics \cite{yannakakis2016psychophysiology, pretty_case_2023}. Examples include measuring heart rate, stress, and cognitive effort, which can be integrated into different game elements to modify the player experience \cite{mandryk2006using, nacke2011biofeedback, robinson2020let}. Besides real-time adaption, studies have tried to understand the relationship between subjective \cite{bowman2018development, pretty2024comparing} and objective metrics \cite{nacke2011biofeedback, pretty2024multimodal} of CL, i.e., whether the user's self-assessed perception of their mental state correlates to its measured biofeedback. \citeauthor{pretty2024multimodal} collected multiple physiological metrics during gameplay (including electroencephalography, electromyography, heart rate, heart rate variability, EDA, and eye blink rate), analysing them against one another \cite{pretty2024multimodal}, and also against subjective measures: the Task Load Index \cite{hancock_NASA_1988} (NASA-TLX) and 
VGDS \cite{bowman2018development} \cite{pretty2024comparing}. Their comparisons indicated strong correlations in some of these pairings, one of which includes EDA, which we aim to reproduce in our study, although under our VR exergame scenario.

Another study investigating hand-eye coordination in a Beat Saber context proposed that exergame training might lead to faster music instrument mastery \cite{rutkowski2021training}. However, that effect is outside the scope of this study, as our experiment aims to combine different research areas in the context of VR. For a comprehensive review of multi-dimensional CL measurements, across a variety of tasks and inclusive of all categories of CL measurement, see~\cite{chen2016robust}.

\subsection{Exergames in VR}
VR offers pedagogical means for users to learn, engage, and enhance their knowledge and skills within a simulated immersive environment. These digital simulations provide a psychological experience dependent on interplays between technology and perceptual processes
\cite{steuer1995defining, takac2023cognitive, renata2024assessment}. They can contribute to a reduction in confounding variables that are typical of laboratory and clinical settings \cite{renata2024assessment}. As attributable to music, VR is a primary medium within entertainment and video games that seeks to provide an engaging experience to the extent of distracting a user from external stimuli \cite{schmidt2018impact}. Extending from this, previous studies have found that situating exergames within a VR context is not only a promising intervention for encouraging physical exercise \cite{huang_exergaming_2020, hedlund2023blocklyvr, yoo2020embedding} but also provides a more immersive experience and feeling of agency than traditional exergames \cite{campo2021immersive}.

In the context of SportsHCI \cite{elvitigala2024sportshci}, several works have explored the exertion aspects of specific sports in an immersive context \cite{gradl2016virtual, sawan2021mixed}, such as jogging \cite{hedlund2023jogging}, climbing \cite{kajastila2016climbingwall}, rowing \cite{delden2020vr4vrt, arndt2018using}, skiing \cite{miura2023exploration}, and cycling \cite{wintersberger2022development}. A significant concern about exertion in VR is attributable to its physically intensive nature, which could lead to discomfort or severe after-effects from prolonged use, such as cybersickness \cite{laviola_discussion_2000}. It is defined as symptoms of eye strain, headache, sweating, disorientation, vertigo, nausea, or vomiting and can occur strictly from visual perception alone \cite{laviola_discussion_2000}. A standardised measure for cybersickness in VR research is the Simulator Sickness Questionnaire (SSQ) \cite{kennedy_simulator_1993}, originally developed to assess simulator sickness in aviators, which will be applied in the current experiment, being in line with current VR experimental protocols for movement-based experiences \cite{hedlund2023blocklyvr, hedlund2023jogging, wintersberger2022development}.

Beyond cybersickness, \citeauthor{arndt2018using} \cite{arndt2018using} used a breathing sensor to gather physiological data during their rowing simulation. In their VR condition, participants showed significantly better results in breaths per minute and rhythm--which are crucial in rowing performance--, when compared to a non-VR baseline. Given that rhythm is highly relevant in our musical exergame scenario, we will also collect physiological data, which is highly interconnected to physical exertion \cite{elvitigala2024sportshci}. However, we will focus on EDA, which has also been linked to CL \cite{pretty2024multimodal}.

\section{Research Aim and Questions}
\label{sec:RQ}

Although aspects of music sophistication and CL have been jointly researched, there are limited studies that attempt to explore this interaction in a VR rhythm-based exergame. As the interaction between music-related activities and executive function activation is well-established in music psychology literature, and working memory is a component of executive function, this could posit CL as a subjective measure of executive function. Furthermore, game difficulty in Beat Saber is referred to as task difficulty to align with terminology used in existing literature. Thus, the present study aims to investigate the interactions of music sophistication, CL, and task accuracy in a VR rhythm exergame as task difficulty increases, given the following research questions (RQ): 
\begin{itemize}
    \item \textbf{RQ1:} What are the effects of task difficulty and game/VR experience on cognitive load? 
    \item \textbf{RQ2:} How does previous music training affect...
    
    \begin{itemize}
    \item \textbf{a)} performance, and
    \item \textbf{b)} cognitive load 
     \end{itemize}
     
     ...in a rhythm exergame?
\end{itemize}

\section{Methodology}
\label{sec:methodology}
\subsection{Participants}
Recruitment of participants was performed via word of mouth, social media websites, 
and advertising posters at the local university. 
Participants were screened relative to equipment safety guidelines\footnote{https://www.meta.com/quest/safety-center/quest-pro/} and demographic characteristics in VR studies \cite{slater_enhancing_2016,laviola2000discussion}. The eligibility criteria were as follows: (1) being 18 to 35 years old; (2) being a near-native level proficient English speaker; (3) not being currently pregnant; (4) having no vision difficulties that cannot be corrected by glasses or contact lenses; (5) having not recently undergone any medical procedure (including cosmetic); (6) not having a pacemaker; (7) not suffering from a heart or serious medical condition; (8) having no physical disabilities that impair balance, motor action, or standing with assistance; and (9) having no history of significant motion sickness, active nausea, and vomiting or epilepsy.
Out of the 74 people who applied and met the screening criteria, 32 participants actively completed the present study. They were given the opportunity to enter a draw for one of two 
gift cards at the completion of the study. The sample consisted of 11 females (34.4\%), 18 males (56.3\%), and 3 non-binary (9.4\%) people aged 18 to 35 years ($M = 23.63, SD = 4.58$). However, the EDA data of five participants was too noisy and the features could not be extracted, making the final number of participants 27. In addition, the MT group split was as such: 13 participants in the low group, and 14 participants in the high group.
\subsection{Materials}
\subsubsection{Demographic measures}
The pre-experimental survey included questions pertaining to participants' previous experience with digital games, VR, and Beat Saber on a 7-point Likert scale, from ``Never played” to ``Play daily” (more details on Table \ref{tab:frequency_statistics}). The questions were as follows: ``What best describes your experience with (1) digital games; (2) Virtual Reality; and (3) Beat Saber?''. 

\paragraph{Gold-MSI}
The Gold-MSI \cite{mullensiefen_musicality_2014} is a 38-item self-report measure of music sophistication comprising five subscales (active engagement, perceptual abilities, MT, singing abilities, and emotions) that contribute to a general music sophistication factor. Out of those, 31 items are scored on a Likert scale ranging from 1 (completely disagree) to 7 (completely agree), with the remaining 7 items being scored under categorical responses (e.g., “I have had formal training in music theory for 0/.5/1/2/3/4-6/7-10/11+ years.”). 
The primary Gold-MSI, derived from 18 items, demonstrated good internal reliability  $\alpha = .88$, with its subscales exhibiting acceptable to good internal reliability between $\alpha = .71$ and $\alpha = .85$ \cite{chee2022musical}, where a higher score signifies higher music sophistication. Responses were scored using the Gold-MSI scoring template\footnote{https://shiny.gold-msi.org/gmsiscorer/} for further analyses. Due to time constraints in this study, the Gold-MSI listening tasks were excluded, and only the derived general and MT scores were analysed. 


\paragraph{Video Game Demand Scale}
The VGDS is a multidimensional instrument designed to assess perceived demand across five domains: cognitive, emotional, controller, exertional, and social~\citep{bowman2018development}. It has demonstrated strong internal consistency (e.g., cognitive: $\alpha = .90$) and good convergent validity with the mental demand subscale in NASA-TLX ($\beta = .927$). The VGDS also shows predictive validity, with each subscale significantly linked to corresponding self-reported effort (e.g., Paas’ mental effort item predicting cognitive demand, $r = .817$). Although we administered the full scale (excluding the social subscale), this study reports on the cognitive subscale as our subjective measure of CL.

\paragraph{Simulator Sickness Questionnaire}
Research in VR warrants the monitoring for simulator sickness or cybersickness symptoms as a necessary duty of care. The SSQ \cite{kennedy_simulator_1993} is a 16-item questionnaire with three subscales (nausea, oculomotor, and disorientation) scored on a 4-point Likert scale from 0 to 3 (0 = none; 1 = slight; 2 = moderate; 3 = severe), and it demonstrates a high reliability of  $\alpha = .91$ \cite{xu_studying_2020}. The SSQ was administered after each VR play session for the researcher to determine the participants' ability to continue with the study.

\subsubsection{Apparatus}
\paragraph{Physiological sensor}
The Emotibit\footnote{https://www.emotibit.com} has 2 Ag/AgCL EDA electrodes for collecting a number of different physiological measurements. It uses an Adafruit Feather M0 WiFi board powered by a 400mAh Lithium ion battery, a microSD card to record the data, and it comes with straps for different placements on the body. In our experiment, it was attached to the participants' right ankle, with them being instructed to keep their feet still as much as possible. The Emotibit Oscilloscope (v1.5.10) was used to stream and record the data, whereas the Emotibit DataParser (v1.11.4) was used to parse the datastream for analysis.

\paragraph{Computer hardware} The MetaQuest Pro\footnote{https://www.meta.com/quest/quest-pro/} was used as a VR head-mounted display due to it being untethered, with its own central processing unit. As shown in Figure \ref{fig:teaser}-left, the surveys and the recording of Beat Saber version 1.31.0 gameplay were completed on separate computers to reduce distraction and ensure the clear separation between task and survey completion, with the recordings being completed using the Open Broadcaster Software\footnote{https://obsproject.com/} (OBS version 29.1.3) saved in MP4 format. 

\paragraph{Game environment} Beat Saber was the game environment used in this experiment. It requires players to react to visual cues of approaching coloured blocks by performing slicing movements to the rhythm or beat of a piece of music. Each block represents a note and it needs to be cut in the direction (horizontally, vertically, or diagonally) indicated by an arrow on each block. Figure \ref{fig:teaser}-right demonstrates the virtual environment and objects players can interact with. The scoring system depends on the swinging angle movement when cutting a note and how precise a cut is through the centre of a note. Since each song has a variable number of notes, and the game difficulty being played highly impacts on the amount of points each cut grants the player, we will use a ratio to normalise the scoring as a metric of accuracy. Hence, we define accuracy as the percentage of the score obtained by the player divided by the maximum possible score in that specific song and difficulty.  
The ranking system in the game is based on this same accuracy, divided in percentiles; however, this is only accessible during gameplay without modifying the game.

Previous studies did not have congruent song choices to reference from. By means of consulting online game forums\footnote{reddit.com/r/beatsaber/comments/rpb8i7/beat\_saber\_music\_packs\_from\_best\_to\_worst/} and extensive playthroughs by the authors, the songs selected needed to be distinguishable in difficulty (see Table \ref{tab:experimental_conditions}). Similar song durations for the three experimental conditions (easy, normal, and hard) were chosen to control for differences in exhaustion beyond task difficulty. For consistent gameplay, the following changes were made to game settings:
\begin{itemize}
    \item Switched on: ``no fail”, ``automatic player height”, fixed note (block) colours (red = right, blue = left), and the game environment was set to ``The First”.
    \item Switched off: ``arc haptics”, ``arc visibility”, ``environment effects” and ``flickering”.
\end{itemize}

\begin{table*}[htbp]
    \centering
    \caption{Beat Saber Song Details for Experimental Conditions\textsuperscript{a} Tutorial, Easy, Normal and Hard}
    \begin{tabular}{lcccc}
        \toprule
        & \textbf{Tutorial} & \textbf{Easy} & \textbf{Normal} & \textbf{Hard} \\
        \midrule
        \textbf{Song Choice} & Balearic Pumping & Rum n Bass & POP/STARS & Natural \\
        \textbf{Song Artist} & Jaroslav Beck & Boom Kitty & K/DA & Imagine Dragons \\
        \textbf{Duration (minutes)} & 02:14 & 03:09 & 03:09 & 03:06 \\
        \textbf{Beats Per Minute} & 111 & 132 & 170 & 100 \\
        \textbf{No. of Notes}\textsuperscript{b} & 170 & 155 & 341 & 553 (526)\textsuperscript{c} \\
        \textbf{Notes per second} & $\sim$1.26 & $\sim$0.82 & $\sim$1.80 & $\sim$2.83 \\
        \textbf{No. of Walls} & 9 & 23 & 41 & 0 \\
        \textbf{No. of Mines} & 6 & 2 & 2 & 0 \\
        \bottomrule
    \end{tabular}
    \label{tab:experimental_conditions}
    \\
    \footnotesize{a) ‘Experimental Conditions’ align with the difficulty levels of the songs played, the only exception being the Tutorial, as participants played a song on ‘Easy’ difficulty level to warm up. b) ‘Notes’ refers to the coloured blocks players interact with in the game. c) Although there are 526 notes in this song, the maximum combo participants could achieve is 553 due to chain notes, making that a more accurate representation for the ‘number of notes’ in this song. }
\end{table*}


\subsection{Procedure}
This study was approved by the Human Research Ethics Committee of the local University. 
Eligible participants provided written informed consent upon arrival to the lab. They completed a survey battery prior to experiencing the VR conditions in the following order (1) demographic data about their previous experience with digital games, VR, and Beat Saber, and (2) the Gold-MSI. Participants were provided a tutorial on how to use the MetaQuest Pro and how to play Beat Saber, which consisted of a tutorial accompanied by an Easy level song to reduce novelty effects for consecutive conditions through gameplay exposure. For each condition, participants were instructed to press ‘play’ after a count of 5 when the researcher said ‘Go’. At the completion of each song, participants reported their scores verbally before removing the headset. These scores were also double-checked by the researchers in the OBS recording after the experiment, and then converted to a percentage over the maximum possible score for that song and difficulty, regarded as accuracy. Participants were then asked to respond to the SSQ and VGDS. After all conditions were done, they would complete a post-experiment survey, which asked participants to rank their preference of difficulty order by level of enjoyment, and the reasoning for their choice.

\subsection{Design}
The present study employed a within-subjects experimental design to evaluate changes to dependent variables across the main independent variable of task difficulty.

\subsection{Data Analysis}

\subsubsection{EDA Pre-Processing}
EDA was recorded at 15 Hz and pre-processed using the \textit{neurokit2} Python package. The raw signal was first smoothed with a fourth-order low-pass Butterworth filter (cutoff: 5 Hz) and then decomposed into tonic and phasic components using the \textit{eda\_process} function, which applies continuous deconvolution. Skin conductance responses (SCRs) were identified using a peak detection algorithm with adaptive thresholding and a minimum amplitude criterion of 0.01 $\mu$S. From this, the following measures were derived for each trial: average tonic EDA, average phasic EDA, number of SCR Peaks, and average SCR amplitude. All EDA variables were standardised ($M = 0, SD = 1$) to enable direct comparison across each other. 

\subsubsection{Statistical Analysis}
All statistical analyses were conducted in R (v4.4.1). Data were screened for missing values, and observations with incomplete data for the variables of interest were excluded, as mentioned above, this was present in five participants' EDA data. Multivariate normality of the EDA variables was assessed separately for each task difficulty condition using Mardia's test. Results showed that assumptions of multivariate normality were not fully met, with only the SCR Peaks variable conforming to normality across conditions and thus SCR Peaks was the only EDA variable kept for the final regression models.

Multiple ANOVA and MANOVA tests were initially conducted to ascertain which variables were suitable for the two linear regression models that aimed to see which objective and subjective data could predict accuracy and subjective CL. Variables were deemed suitable if they could display significant (or near-significant) differences across the 3 difficulties. A one-way MANOVA was conducted with task difficulty (easy, normal, and hard) as the independent variable, and four standardised EDA indices: tonic EDA, phasic EDA, SCR Peaks, and SCR amplitude as dependent variables. A follow-up ANOVA and post hoc tests were conducted for variables showing significant effects. Subjective CL, as measured by the cognitive subscale of the VGDS, was also compared across difficulty conditions using a one-way ANOVA, with follow-up post hoc tests. This was also done for comparing accuracy across difficulties. Additional regression analyses were performed to examine the predictive value of prior experience variables (Beat Saber, VR, and digital game experience) on accuracy, subjective CL, and SCR Peaks.

Finally, to address the research questions of this study, two multimodal linear regression models were conducted to evaluate the predictive value of task difficulty, Beat Saber experience, VR experience, digital games experience, MT group, and SCR Peaks on subjective CL and accuracy (with the addition of subjective CL to the latter model). Interaction effects were tested for the particular variables of interest (MT group \& CL, and MT group \& difficulty), however none were significant, and thus the final models only included main effects. This has added benefit of reducing multicollinearity, as well as simplifying the models. 


\begin{figure}[h]
  \centering
  \includegraphics[width=\linewidth]{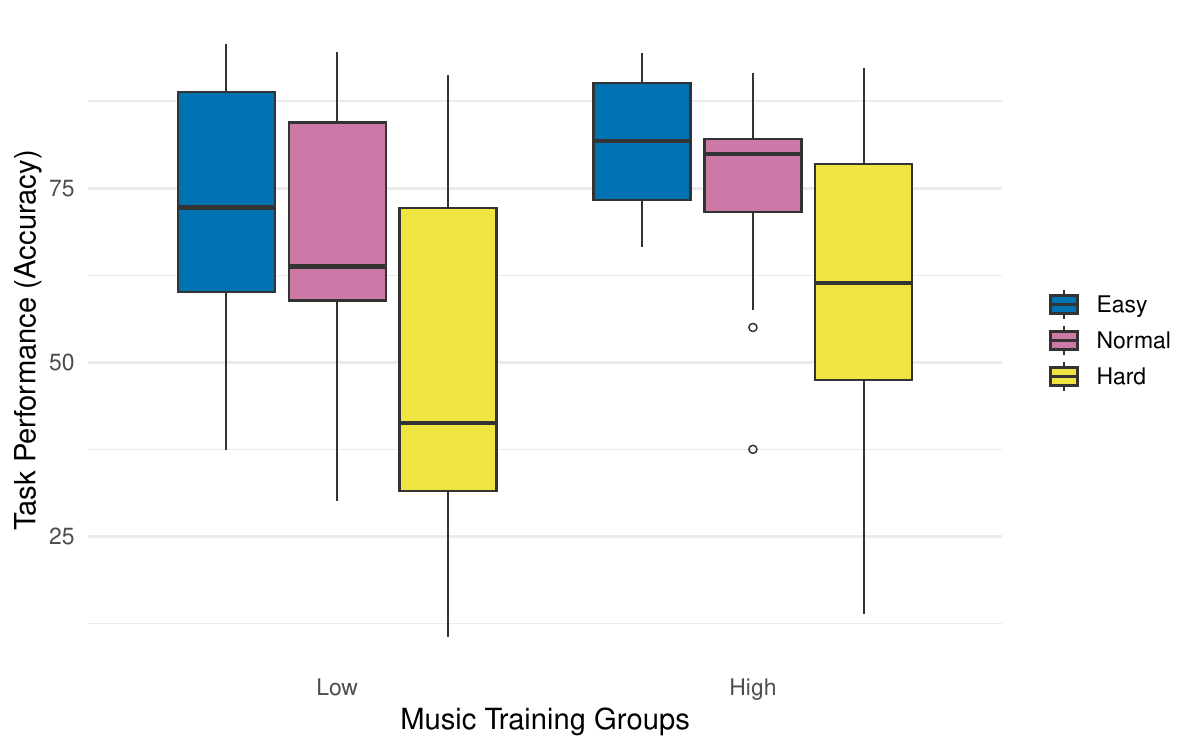}
  \caption{Changes in Task Performance (Accuracy) between Musical Training groups as Task Difficulty increases.}
  \Description{Changes in Task Performance between Musical Training groups as Task Difficulty increases.}
  \label{fig:taskperf}
\end{figure}

\section{Results}
\label{sec:results}
\subsection{Demographics}
Frequencies of responses to pre-experimental survey questions can be found on Table \ref{tab:frequency_statistics}, which describes players’ experience with digital games, VR, and Beat Saber on a 7-point Likert scale. On average, 78.1\% participants played digital games once a week or more ($M = 6.22, SD = 1.04$), 84.4\% have played in VR at least once a year ($M = 3.69, SD = 1.45$), and 62.5\% have played Beat Saber at least once in the last 5 years ($M = 2.91, SD = 1.59$). An independent samples t-test found no statistically significant difference between male ($M = 72.78, SD = 4.48$) and female ($M = 69.18, SD = 4.02$) participants for general musical sophistication scores ($p = .588$). Participants' mean SSQ scores were within the expected range across all conditions: easy ($M = 8.32, SD = 2.51$), normal ($M = 12.06, SD = 3.24$) and hard ($M = 16.29, SD = 4.57$). Two participants scored above the standard threshold and were monitored to determine their fitness to continue the study.


\begin{table}[htbp]
    \centering
    \caption{Frequency Statistics of Participants' (N=32) Experience with Digital Games, VR, and Beat Saber}
    {%
    \begin{tabular}{llcccc}
        \toprule
        \textbf{Measure} & \textbf{Score} & \textbf{ƒ} & \textbf{Relative ƒ} & \textbf{cƒ} & \textbf{Perc.} \\
        \midrule
        \multirow{7}{*}{\textbf{\small Dig. Games experience}} 
        & 7 & 17 & 53.13\% & 32 & 100 \\
        & 6 & 8 & 25.00\% & 15 & 46.88 \\
        & 5 & 5 & 15.63\% & 7 & 21.88 \\
        & 4 & 1 & 3.13\% & 2 & 6.25 \\
        & 3 & 1 & 3.13\% & 1 & 3.13 \\
        & 2 & 0 & 0.00\% & 0 & 0.00 \\
        & 1 & 0 & 0.00\% & 0 & 0.00 \\
        \midrule
        \multirow{7}{*}{\textbf{ \small VR experience}} 
        & 7 & 1 & 3.13\% & 32 & 100 \\
        & 6 & 3 & 9.38\% & 31 & 96.88 \\
        & 5 & 3 & 9.38\% & 28 & 87.50 \\
        & 4 & 11 & 34.38\% & 25 & 78.13 \\
        & 3 & 9 & 28.13\% & 14 & 43.75 \\
        & 2 & 2 & 6.25\% & 5 & 15.63 \\
        & 1 & 3 & 9.38\% & 3 & 9.38 \\
        \midrule
        \multirow{7}{*}{\textbf{\small Beat Saber experience}} 
        & 7 & 0 & 0.00\% & 32 & 100 \\
        & 6 & 2 & 6.25\% & 32 & 100 \\
        & 5 & 1 & 3.13\% & 30 & 93.75 \\
        & 4 & 12 & 37.50\% & 29 & 90.63 \\
        & 3 & 5 & 15.63\% & 17 & 53.13 \\
        & 2 & 1 & 3.13\% & 12 & 37.50 \\
        & 1 & 11 & 34.38\% & 11 & 34.38 \\
        \bottomrule
    \end{tabular}%
    }
    \label{tab:frequency_statistics}
    \newline
    \footnotesize{1 = “Never played”; 2 = “Not played in the last 5 years”; 3 = “Played at least once in the last 5 years”; \\4 = “Play at minimum once a year”; 5 = “Play at minimum once a month”; 6 = “Play at minimum once a week”; 7 = “Play daily”}.
\end{table}

\subsection{Qualitative Results}
Following the main experiment, participants ranked the difficulty levels by which they found the most enjoyable accompanied by reasoning. Of interest was their highest ranked difficulty, which was hard ($56\%, n = 18$), normal ($31\%, n = 10$), and easy ($13\%, n = 4$). Feedback from the hard condition was clustered thematically: the need to concentrate (“the harder it was, the more I had to actually focus”), high stimulation (“it was a fun and novel challenge”) and flow (“appropriate difficulty for my skill level”; “…helped me to get into the zone”). The normal difficulty was described as “the perfect balance for my current expertise in the game”, in addition to it being the “right amount of difficulty for the game”. Participants who enjoyed the easy difficulty the most emphasised on performance, stating it was “easier to slash the cubes, resulting in [a] higher score and better ranking”.

\subsection{Multivariate Analysis of Variance of EDA by Difficulty}

A MANOVA with task difficulty as the predictor revealed a non-significant multivariate effect on the combined EDA indices (Pillai's Trace $= .137, F(8, 138) = 1.27, p = .267$). Follow-up univariate ANOVAs indicated that only SCR Peaks differed significantly by task difficulty ($F(2, 71) = 3.54, p = .034$), with an almost marginal effect also found for SCR Amplitude ($F(2, 71) = 2.95, p = .059$). Post hoc comparisons (Tukey HSD) for SCR Peaks indicated a significant difference between easy and hard tasks ($p = .026$), with more SCR Peaks during hard tasks (see Table \ref{tab:posthoc_tukey}-top). No significant differences were found between normal and either easy or hard conditions. Based on this and the previous normality results, SCR Peaks was chosen for both of the final regression models.

\subsection{Effects of Difficulty on Accuracy and Subjective CL}

A one-way ANOVA revealed a significant effect of task difficulty on subjective CL ($F(2, 93) = 18.62, p < .001$). Post hoc Tukey HSD tests confirmed that participants reported significantly higher CL in hard tasks compared to both normal ($p = .002$) and easy tasks ($p < .001$), and also reported significantly higher CL in normal compared to easy tasks ($p = .028$), as seen in Table \ref{tab:posthoc_tukey}-bottom. Assumptions of normality (Shapiro-Wilk $p = .077$) and homogeneity of variance (Levene's $F(2, 93) = 2.11, p = .128$) were met. As these findings indicate, the songs were different in difficulty, and thus the difficulty variable was included in the final regressions.

A one-way ANOVA revealed a significant effect of task difficulty on accuracy ($F(2, 93) = 10.88, p < .001$). Post hoc Tukey HSD tests indicated that accuracy was significantly lower in hard tasks compared to both easy ($p < .001$) and normal tasks ($p = .003$), while no significant difference was observed between easy and normal tasks ($p = .499$). The assumption of normality was met (Shapiro-Wilk $p = .099$), but Levene’s test indicated a violation of the homogeneity of variance assumption ($F(2, 93) = 4.78, p = .011$). A non-parametric Kruskal-Wallis test confirmed the effect of task difficulty on accuracy ($\chi^2(2) = 14.51, p < .001$). Pairwise Wilcoxon tests further supported the finding that accuracy was significantly lower in hard compared to easy ($p < .001$) and normal tasks ($p = .025$), with no significant difference between easy and normal tasks ($p = .857$), as seen in Table \ref{tab:posthoc_wilcoxon}. 

\begin{table}[ht]
\centering
\caption{Post hoc comparisons for SCR Peaks and Cognitive Load (Tukey HSD)}
\begin{tabular}{lccc}
\toprule
Comparison & Mean Difference & 95\% CI & $p$ \\
\midrule
\multicolumn{4}{l}{\textbf{SCR Peaks}} \\
Hard vs Easy & 0.73 & [0.07, 1.38] & .026* \\
Normal vs Easy & 0.33 & [-0.33, 0.99] & .460 \\
Normal vs Hard & -0.40 & [-1.06, 0.26] & .329 \\
\multicolumn{4}{l}{\textbf{Cognitive Load}} \\
Hard vs Easy & 190.10 & [115.67, 264.54] & <.001*** \\
Normal vs Easy & 81.77  & [7.34, 156.20] & .028* \\
Normal vs Hard & -108.33 & [-182.77, -33.90] & .002** \\
\bottomrule
\end{tabular}
\label{tab:posthoc_tukey}
\end{table}

\begin{table}[ht]
\centering
\caption{Post hoc comparisons for Accuracy (Wilcoxon pairwise tests)}
\begin{tabular}{lccc}
\toprule
Comparison & Med. Diff. & Group Medians (IQR) & $p$ \\
\midrule
Hard vs Easy & -22.0 & 53.8 (41.5) vs 75.8 (20.4) & <.001*** \\
Hard vs Normal & -24.6 & 53.8 (41.5) vs 78.4 (22.6) & .025* \\
Normal vs Easy & 2.6 & 78.4 (22.6) vs 75.8 (20.4) & .857 \\
\bottomrule
\end{tabular}
\label{tab:posthoc_wilcoxon}
\end{table}

\subsection{Experience Factors as predictors of accuracy, subjective CL, and SCR Peaks}

A linear regression assessed whether Digital Games, VR, and Beat Saber experience predicted subjective CL. The model was significant ($F(3, 92) = 4.81, p = .004$), explaining 10.7\% of the variance ($R^2_{adj} = .107$). Both Digital Games ($\beta = -35.09, p = .013$) and Beat Saber experience ($\beta = -32.91, p = .016$) were significant negative predictors, while VR experience was not ($p = .22$). Multicollinearity was low (all $VIFs <3$), and visual inspection of histograms and Q-Q plots displayed normal residuals. 

Additionally, a linear regression assessed whether Digital Games, VR, and Beat Saber experience predicted accuracy. The model was significant ($F(3, 92) = 9.91, p < .001$), explaining 21.9\% of the variance ($R^2_{adj} = .220$). Digital Games ($\beta = 5.42, p = .006$) and Beat Saber experience ($\beta = 4.08, p = .030$) were significant positive predictors of accuracy, while VR experience was not ($p = .39$). Multicollinearity was low (all $VIFs <3$), and visual inspection of histograms and Q-Q plots displayed normal residuals. As such, both Digital Games and Beat Saber experience were significant in these regressions, thus they were included in the final regression models.

A separate regression predicting SCR Peaks from the same experience variables, as well as MT group, was not significant ($F(3, 70) = 0.63, p = .595, R^2_{adj} = -0.015$). The individual experience factors did not meaningfully predict EDA response, however MT group did ($\beta = -0.54, p = .028$). Multicollinearity was low (all $VIFs <3$), and visual inspection of histograms and Q-Q plots displayed normal residuals. 

\subsection{Combined Predictive Models of Accuracy and Subjective Cognitive Load}

\begin{figure}[h]
  \centering
  \includegraphics[width=\linewidth]{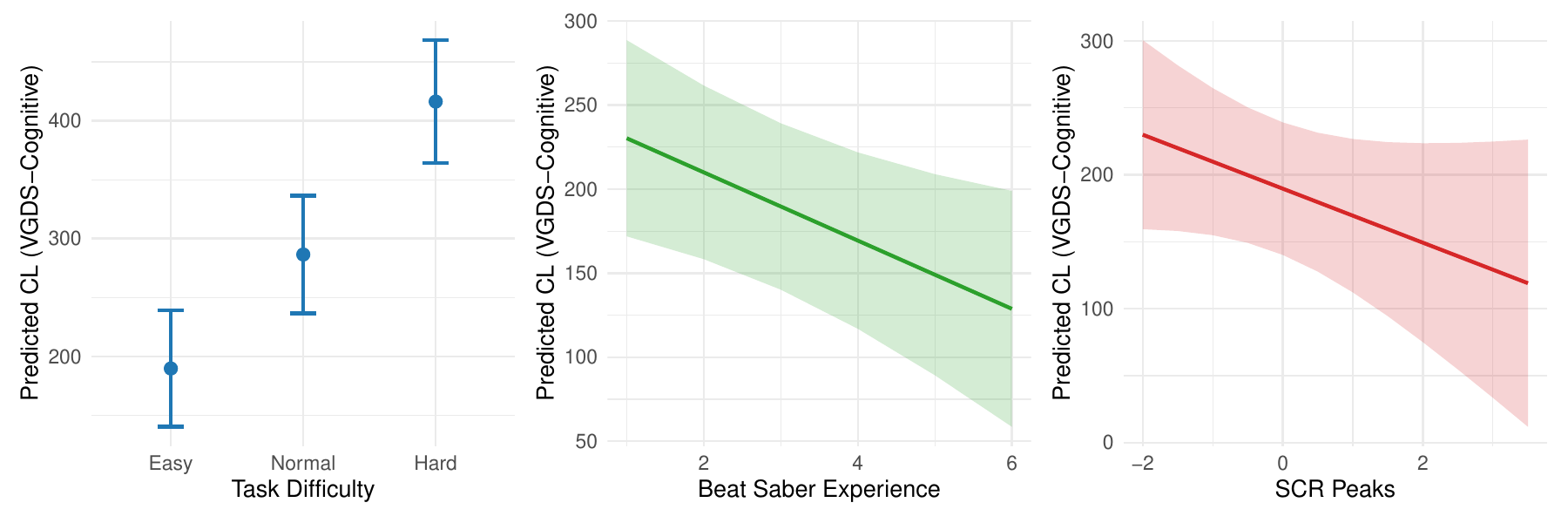}
  \caption{Predicting Subjective Cognitive Load from Task Difficulty, Beat Saber experience, and SCR Peaks}
  \Description{Predicting Subjective Cognitive Load from Task Difficulty, Beat Saber experience, and SCR Peaks}
  \label{fig:CLpredictors}
\end{figure}

The final regression model to predict subjective CL including task difficulty, Digital Games experience, Beat Saber experience, MT group, and SCR Peaks (see Table \ref{tab:regression}-top) was also significant, $F(6, 67) = 12.43, p < .001$, explaining 48\% of the variance ($R^2_{adj} = .48$). Significant predictors included task difficulty (both normal ($\beta = 96.80, p = .002$) and hard ($\beta = 226.66, p < .001$) vs. easy), Beat Saber experience ($\beta = -25.42, p = .003$), Digital Games experience ($\beta = -28.56, p = .016$), but not SCR Peaks ($\beta = -20.17, p = .133$) or MT group ($\beta = 21.27, p = .42$). Normality testing of residuals showed normal histogram and Q-Q plots, as well as low multicollinearity (all $VIFs < 2$), although the Shapiro-Wilk test indicated non-normal residuals ($p = .024$). For visual representation of this model, see Figure~\ref{fig:CLpredictors}.

\begin{figure}[h]
  \centering
  \includegraphics[width=\linewidth]{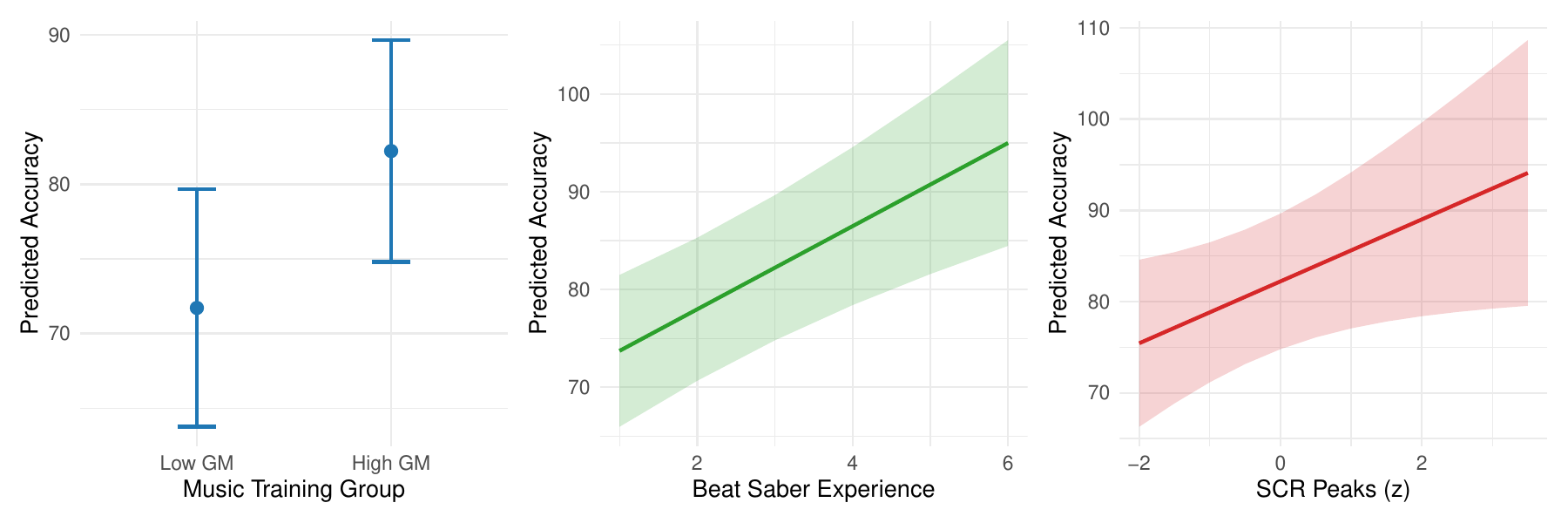}
  \caption{Predicting Accuracy from MT Group, Beat Saber experience, and SCR Peaks}
  \Description{Predicting Accuracy from MT Group, Beat Saber experience, and SCR Peaks}
  \label{fig:perfPredictors}
\end{figure}

A final regression model was conducted to predict task accuracy from subjective CL, music training group, task difficulty, Digital Games experience, Beat Saber experience, and SCR Peaks, as seen in Table \ref{tab:regression}-bottom. The overall model was significant, $F(7, 66) = 20.50, p < .001$, explaining 65.2\% of the variance in accuracy ($R^2_{adj} = .652$). Subjective CL was a significant negative predictor ($\beta = -0.053, p < .001$), indicating that higher perceived CL was associated with lower performance. Music training group significantly predicted accuracy ($\beta = 10.51, p = .002$), with participants in the high music training group achieving higher scores overall. Difficulty also influenced accuracy, with significantly lower performance on hard tasks compared to easy ($\beta = -12.27, p = .018$), but no significant difference between normal and easy conditions. Both Digital Games experience ($\beta = 3.72, p = .014$) and Beat Saber experience ($\beta = 4.26, p < .001$) were significant positive predictors of accuracy, and SCR Peaks also contributed modestly to the model ($\beta = 3.39, p = .043$). Residuals were approximately normal based on visual inspection of Q-Q plots and histograms, and multicollinearity was low (all $VIFs < 2$). For a visualisation of the predictors of accuracy, see Figure~\ref{fig:perfPredictors}.

\begin{table}[ht]
\centering
\caption{Regression coefficients for predictors of Cognitive Load, SCR Peaks, and Accuracy}
\begin{tabular}{lccc}
\toprule
Predictor & $\beta$ & SE & $p$\\
\midrule
\multicolumn{4}{l}{\textbf{Final Combined Model Predicting Cognitive Load}} \\
Difficulty (Hard) & 226.66 & 32.03 & <.001***\\
Difficulty (Normal) & 96.80 & 31.07 & .002** \\
Dig. Games experience & -28.56 & 11.59 & .016* \\
Beat Saber experience & -25.42 & 8.10  & .003**  \\
SCR Peaks & -20.17  & 13.30 & .133  \\
MT Group & 21.27 & 26.18 & .420 \\
\multicolumn{4}{l}{\textbf{Final Combined Model Predicting Accuracy}} \\
Subjective CL & -0.053 & 0.01 & <.001***  \\
MT Group & 10.51 & 3.32 & .002**  \\
Difficulty (Hard) & -12.27 & 5.10 & .018* \\
Difficulty (Normal) & -3.81 & 4.88 & .438  \\
Dig. Games experience & 3.72 & 1.48 & .014*  \\
Beat Saber experience & 4.26 & 1.07 & <.001***  \\
SCR Peaks & 711.38 & 346.72 & .043*  \\
\bottomrule
\end{tabular}
\label{tab:regression}
\end{table}

\section{Discussion}
\label{sec:discussion}

The present study investigated whether musical training could explain variations in CL and task accuracy in the VR rhythm game Beat Saber as task difficulty increased. Through regression analyses, we determined that while task difficulty and gaming experience significantly predicted subjective CL, musical training alone did not significantly contribute to subjective perceptions of cognitive effort. However, musical training significantly predicted greater accuracy in gameplay. These results imply that musical expertise contributes specifically to improved performance, potentially via enhanced motor coordination or visual-spatial processing, rather than lowering perceived cognitive demands.

\subsection{Task Difficulty, Experience, and Cognitive Load}
Considering \textbf{RQ1}, our regression analyses revealed that task difficulty significantly impacted subjective CL, with the hard condition eliciting the highest perceived cognitive effort, followed by normal and easy difficulties. Importantly, specific Beat Saber experience and broader Digital Games experience significantly predicted lower subjective CL, suggesting that familiarity with the game mechanics and digital interaction environments effectively reduced perceived cognitive demands. In contrast, musical training did not significantly predict subjective CL, highlighting that task-specific and domain-specific experiences rather than broader musical expertise influenced participants' subjective experience of CL.

These findings underscore the critical role of domain-specific familiarity in managing CL within immersive and interactive digital environments. Furthermore, physiological measures (SCR Peaks) provided additional context, as increased arousal correlated with better task accuracy, reinforcing the concept that moderate physiological activation may facilitate task engagement and concentration. Overall, this reinforces CL theory's assertion that familiarity and expertise significantly mediate perceived difficulty, particularly in complex, multimodal tasks such as an immersive audio-visual exergame.

Although these results underline CL as a key differentiator across task conditions, potential confounding effects from VR novelty should be acknowledged. Considering our participant sample consisted of limited previous VR experience, and despite instructions to report CL based on the task alone, CL results can be conflated by way of exposure to a virtual environment \cite{anglin2017visuomotor, frederiksen2020cognitive}. 

\subsection{Music Training \& Cognitive Load}

As for \textbf{RQ2a}, MT group was not a significant predictor of subjective CL in the regression model. In contrast, task difficulty, Beat Saber, and Digital Games experience significantly predicted subjective CL, with participants reporting higher load in more difficult conditions and lower load if they had greater familiarity with the game and games generally. This suggests that perceived mental effort was more strongly shaped by the nature of the task and prior experiences than by MT (see Figure~\ref{fig:CLpredictors}). These results suggest that, within this context, MT does not generalise to perceived mental demand, however, one regression found that expertise did predict objective CL. Thus, their inclusion helped account for individual variation, which may interact with task perception in more complex or indirect ways. This incongruence of findings highlights the influence of domain-specific expertise, such as regular exposure to game mechanics, on perceived cognitive demand. That is, rather than being modulated by broader cognitive skills associated with MT, subjective CL in this task appears more tightly linked to situational familiarity, whereas objective CL is linked more specifically to task-specific difficulty~\cite{sweller_cognitive_2011}.

These findings also did not replicate previous results reported by Pretty et al. \cite{pretty2024multimodal}, who found stronger associations between subjective and physiological measures of CL. The discrepancy may be attributed to differences in task complexity, participant familiarity with VR, the distinct physiological measures used, or the nature of cognitive demands posed by Beat Saber relative to the tasks employed in previous studies. Further research should aim to clarify these methodological variations and examine under what conditions different CL measurements converge.


\subsection{Music Training \& Task Accuracy}
In the final regression model predicting accuracy, related to \textbf{RQ2b}, MT group emerged as a significant predictor, alongside task difficulty, SCR Peaks, Beat Saber and Digital Games experience. Participants with higher MT performed more accurately overall, even when accounting for other factors. This aligns with existing research suggesting that MT supports visual-spatial processing \cite{strong2019cognitive}, a skill essential in Beat Saber’s fast-paced virtual environment \cite{tian_impact_2023}.

Task difficulty and game experience also significantly influenced accuracy — harder tasks reduced accuracy, while both specific and general game familiarity improved it. Although no significant interaction or moderating factor was found between expertise and subjective CL on task performance, the result of music training and CL predicting accuracy does corroborate the findings in CL literature, despite no interactions being found~\cite{sweller_cognitive_2011}. Additionally, higher SCR Peak values were associated with better accuracy, possibly indicating that physiological arousal plays a facilitating role in focused, time-sensitive gameplay (see Figure~\ref{fig:perfPredictors}), though it may just be reflective, rather than predictive, of the increased mental focus of the harder difficulties.


Visual inspection of accuracy data (Figure~\ref{fig:taskperf}) suggests that although the higher MT group exhibited accuracy levels closer to each other, the low MT group had a wider spread of accuracy across all difficulty conditions notwithstanding lower medians. This addresses a critical pitfall of the musicians versus non-musicians dichotomy within the literature, which \citeauthor{talamini_musicians_2017} \cite{talamini_musicians_2017} underscore to be from expertise acquired via informal training, as opposed to formal, and the ignorance of other traits that contribute towards musical abilities. A possible explanation for this observation in our results is the grouping method for low and high MT. These findings support the need for more holistic classifications of musical expertise \cite{baker_examining_2020, dahary_relationship_2020, matziorinis_is_2023}, and contribute to the broader discussion around the relationship between MT and executive function \cite{bowmer2018investigating}.


\subsection{Limitations \& Future Directions}
The pilot nature of this study imposes inherent limitations for real-world applications beyond providing insight into interactions between CL, task difficulty, and expertise. Notably, the small sample size for music research and the method of categorising groups. Firstly, the sample may be biased due to recruiting on social media and within a university context, 
resulting in a younger sample than in previous studies. However, this variates from the literature, which is known for observing CL in young children or older adult populations within educational research \cite{yeung1998cognitive, berends2009effect, chaddock-heyman_musical_2021}). Secondly, MT groups were categorised by a singular item, as opposed to multiple or an entire subscale, which is a deviation from previous research \cite{dahary_relationship_2020, matziorinis_is_2023}. In supplement to this, sociodemographic factors such as occupation and income were not assessed despite eliciting a high impact on music sophistication collectively in the original Gold-MSI study \cite{mullensiefen_musicality_2014}. Furthermore, future research would benefit from including objective listening tasks, as participants may underestimate or overestimate their responses on the Gold-MSI survey battery, potentially skewing results. It is recommended for future work using the Gold-MSI to account for sociodemographic and objective listening task differences to better compare participant characteristics between studies.
Although not assessed, additional insight could be derived from observing flow \cite{csikszentmihalyi2000beyond}, given that qualitative statements from participants infer different levels of challenge being appropriate for their current skillset. Furthermore, in observing subjective CL within a VR rhythm game task, there remains uncertainty in how much participants’ performance influenced their reporting. Future research is encouraged to explore the relationship between knowledge of performance and subsequent subjective CL, in addition to examining the effects of each VGDS item and accounting for self-efficacy regarding task performance. 

\section{Conclusion}
\label{sec:conclusion}

In summary, this study attempted to explore the assumption that musical expertise could explain the accumulation of CL and task accuracy in the VR rhythm game Beat Saber across different task difficulties. This investigation found that changes in task difficulty led to an increase in CL and a decrease in accuracy. While MT was not a significant predictor of subjective CL, it was a significant predictor of accuracy. These findings indicate that although MT may not influence perceived mental effort, it can contribute to more accurate gameplay, possibly due to enhanced visual-spatial processing or motor coordination. From a theoretical viewpoint, this study adds to CL literature by meeting assumptions of element interactivity through a simple yet demanding commercial game environment. Moreover, the findings support prior research suggesting that domain-specific experience, in our context digital games and Beat Saber, is linked to improved accuracy and reduced perceived CL. Future research would benefit from measuring constructs such as flow to provide further insight, as qualitative responses pointed to their relevance in shaping user experience. Thus, the present study offers an initial basis for understanding how task difficulty, game experience, and individual differences contribute to variability in CL and task accuracy, and suggests that music expertise effects may require more refined classification to fully capture their impact in VR rhythm exergames.

\bibliographystyle{ACM-Reference-Format}
\bibliography{sample-base}

\end{document}